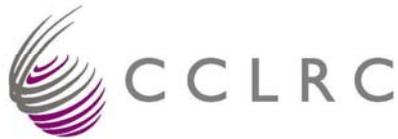 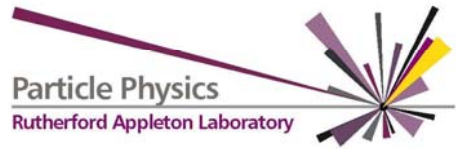

# The Top Quark – 2006 and Beyond

**John Womersley**

*Director of Particle Physics, CCLRC Rutherford Appleton Laboratory*
*Chilton, Didcot, OX11 0QX, United Kingdom*
*E-mail:* `j.womersley@rl.ac.uk`

We know there is new physics at the electroweak scale, but we don't know what it is. Right now, the top quark is our only window on to this physics. In almost all models of electroweak symmetry breaking, top either couples strongly to new particles or its properties are modified in some way. Top is being studied in detail at the Fermilab Tevatron. Its production cross section has been measured in a variety of channels; its mass has been determined to better than 2%, and can be used to constrain the mass of the Higgs. Top quark decays have been tested and non-standard production mechanisms searched for. Single top production probes the electroweak properties of top, and has not yet been observed; searches are now closing in on this process and it should be seen soon. So far, all of the top quark's properties are consistent with the Standard Model. However, the data still to come at the Tevatron will increase the precision of all these measurements, and the enormous statistics available at the LHC will open up new possibilities such as observation of spin correlations and perhaps even CP violation in the top sector.

**Introduction**

What is the universe made of? This is a very old question, and one that has been approached in many ways. The only reliable way to answer this question is by directly enquiring of nature, through experiments. Decades of experimentation with accelerators, and theoretical synthesis of the results, have culminated in what we call the "Standard Model" – a theory of matter and forces, specifically a quantum field theory describing point-like fermions (quarks and leptons) which interact by the exchange of vector bosons (photons, $W^{\pm}$ and $Z$, and gluons). If all we were talking about was one of six quarks in the standard model, I doubt that we'd be at this workshop. We are here in Portugal because we know *a revolução está vindo* – the revolution is coming. The standard model makes precise and accurate predictions, and provides an understanding of what nucleons, atoms, stars, you and me are made of. Nonetheless, its spectacular success in describing phenomena at energy scales below 1 TeV is based on at least one unobserved ingredient – the SM Higgs – whose mass is unstable to loop corrections, requiring something like supersymmetry to solve, and which has an energy density $10^{60}$ times too great to exist in the universe we live in. The way forward is through experiment (and only experiment): this is both tantalizing – we know the answers are accessible – and also a bit frustrating – since we have known this for 20 years.

Meanwhile, back in the universe, we now know there is much more mass than we'd expect from the stars we see, or from the amount of helium formed in the early universe: there is far more non-baryonic dark matter than there is quarks and leptons. Moreover the velocity of distant galaxies shows there is some kind of energy driving the expansion of the universe, as well as mass slowing it down, which we call dark energy. Together, this means we do not know what 96% of the universe is made of. Intriguingly, there are arguments that dark matter, at least, may be related to electroweak scale physics and thus accessible in accelerator experiments.

What does any of this have to do with top? We know there's new physics at the electroweak scale, but we really don't know what it is. Right now, the top quark is our only window on this physics. Top couples strongly to the Higgs field: what is this telling us? Top offers a window on fermion mass generation: does it really happen through Higgs Yukawa couplings? Top provides a unique physics laboratory at the intersection of QCD (its production), electroweak physics (its decay) and Higgs or other new physics.

**Top and new physics**

Almost all solutions to electroweak symmetry breaking have a connection with the top quark[1]. In supersymmetry, the top Yukawa coupling is modified with respect to its SM value, the mass scale of top partners must be low (which is not true of other superpartners), and new physics associated with top may be the first to be seen. In Little Higgs models, there is a new vector-like top-partner T, with a mass ~ 1–2 TeV, which mixes with top, and decays to th, tZ, and bW.

In strongly-coupled models like Technicolor and its descendents in which mass is dynamically generated, top is special because of its large mass: this requires extra interactions (like topcolor) and produces resonances in tt, tb (which would be seen in single top production in the s-channel). In models with modified spacetime such as extra dimensions, there is not such a special role for top, but one can have nonstandard $t\bar{t}$ production through KK resonances.

**Tevatron Status**

The world's only source of top quarks is the Fermilab Tevatron collider. The accelerator is performing well, and roughly 1.4 fb$^{-1}$ has been delivered to each of the CDF and DØ experiments. Electron cooling is installed and working in the recycler ring, which will increase the luminosity by a factor of roughly two; we can expect 2 fb$^{-1}$ by autumn 2006, 4 fb$^{-1}$ by autumn 2007, and 8 fb$^{-1}$ by the end of the run (Autumn 2009).

**Top identification**

In the standard model, the decay t → Wb dominates, and the final states of $t\bar{t}$ production are therefore simply defined by the way in which the two W bosons decayed. In 30% of tt events, an electron/muon plus jets and missing $E_T$ are produced; in about 5% there are either two electrons, two muons or an electron plus a muon, together with jets and missing $E_T$. The remainder yield all-hadronic final states or final states with one or more taus.

Top therefore requires an excellent understanding of the whole detector and of QCD – the performance of triggering, tracking, b-tagging, and electron, muon, jet, and missing $E_T$ identification and measurement must all be understood and modelled[2]. A big effort to understand Jet Energy Scale is needed for event kinematics and top quark mass, b-tagging is required both to reduce backgrounds and to reduce combinatorics in measurements of top quark properties, and sophisticated analysis techniques can maximise sensitivity to rare processes and to deviations from the standard model. All this takes teamwork and efficient tools.

The jet energy scale is the dominant uncertainty in many measurements of the top quark[3]. CDF and DØ use different approaches to determine the jet energy scale and uncertainty. In CDF, the jet scale is taken mainly from single particle response convoluted with a jet fragmentation model, and cross-checked with photon/Z-jet $p_T$ balancing. The uncertainty in Run II is roughly 3% and further improvements are in progress. In DØ, the jet scale comes mainly from photon-jet $p_T$ balance and is cross-checked with the closure tests in photon/Z+jet events. A new Run II calibration (with an uncertainty of roughly 2%) will come out soon. Recently, in-situ calibration with hadronic W decays has been successfully used by both collaborations to reduce the jet energy scale uncertainties in top mass measurements. We can expect results on the b-jet energy scale from photon/b-jet $p_T$ balance and Z → $b\bar{b}$ soon.

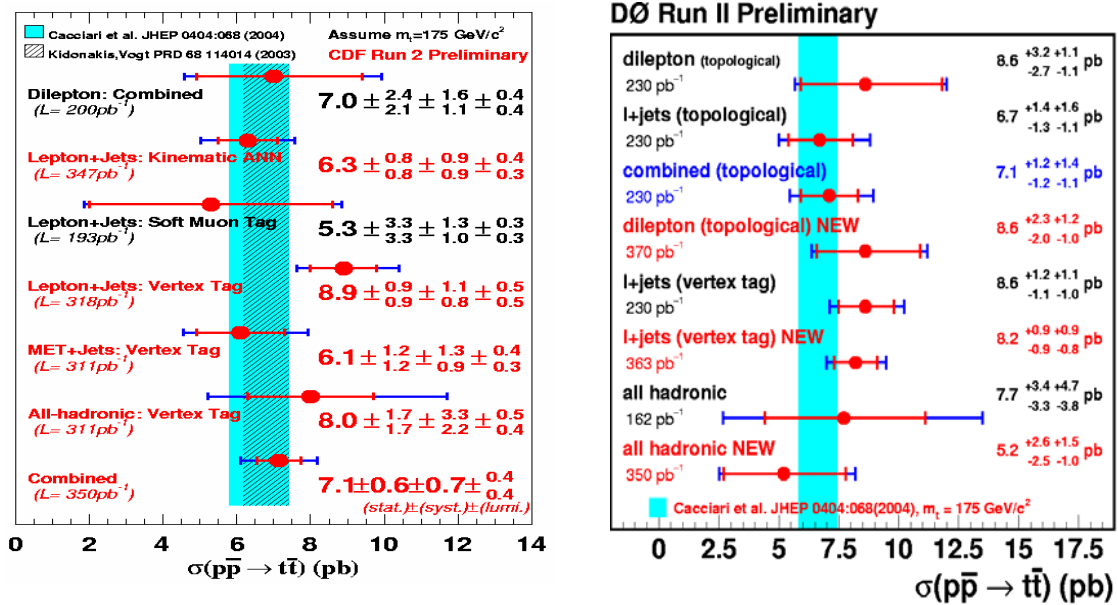

**Figure 1 - Top cross section measurements in various channels from CDF (left) and DØ (right)**

A significant body of experience has been gained in b-tagging at the Tevatron experiments[4]. Both have developed multiple b-tagging tools. Many issues deserve attention for the LHC: the alignment of the silicon tracking detector, understanding of charge deposition, understanding the material in the tracking volume, tracking simulation and its relation to reality, Monte Carlo scale factors, and the determination of efficiencies from data – calibration data must be collected at appropriate $E_T$ and $\eta$.

There has been significant progress on event generators in the past few years[5]. Recent innovations include top-quark production with spin correlations; single top production including $2 \to 2$ and $2 \to 3$ processes with proper matching; tree level generators with additional multi-jets in the final state; prescriptions to match tree-level and showering Monte Carlos without double counting; generators with full NLO corrections to top production processes; improvements to b-quark fragmentation[6]; and inclusion of top production and decays due to interactions beyond the Standard Model. Unfortunately, no single generator incorporates all the desired features, so the experiments need to be aware of the strengths and weaknesses of each.

**Top production**

If the top is "just" a very heavy quark, its production cross section can be calculated in QCD. Cross section measurements can be made in a variety of channels and tested for consistency[7]. Dileptons offer the cleanest channel, while lepton plus jets have higher yields. Figure 1 shows the cross sections measured by CDF and DØ. They are consistent with each other and with QCD. There is an ongoing effort to combine measurements within and among experiments.

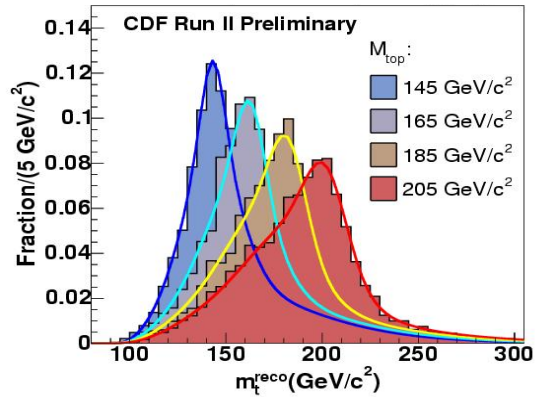

**Figure 2 - Example of templates corresponding to different generated top quark masses.**

**Extracting the top mass**

Lepton plus jets events have traditionally yielded the most precise determination of the top quark mass. There are two basic techniques to extract the top quark mass from a top-rich data sample[8]. In the "template method," a quantity is extracted from each event, for example a reconstructed top mass, and the best fit is found for the distribution of this quantity to a series of Monte Carlo "templates" corresponding to different generated top quark masses (see Figure 2). In the "matrix element" (or "dynamic likelihood") method, one calculates a likelihood distribution from each event as a function of hypothesised top mass, and then multiplies these distributions to get the overall likelihood.

Both experiments are now simultaneously calibrating the jet energy scale in situ using the W → jj decay within top events. A combined fit to the top mass and to an overall scale parameter for the jet energy scale is made, using the known W mass. Figure 3 shows an example from DØ where the jet scale is shifted by about 3%. The technique reduces the impact of jet energy scale uncertainties on the top mass but it cannot completely eliminate them, since it provides no information on $E_T$ or $\eta$ dependence of the scale, or on differences in scale between b-jets and light quarks.

Dilepton events[9] offer a purer sample and reduced sensitivity to the jet energy scale, at a cost of reduced statistics and the inability to incorporate a W → jj calibration. With more statistics, we expect the dilepton channels to become increasingly competitive with the lepton + jets channel.

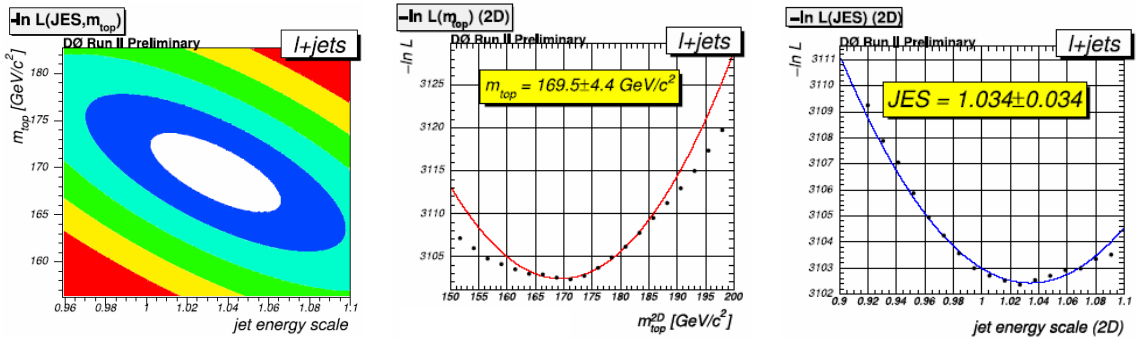

**Figure 3 - Example of simultaneous fit to top mass and jet energy scale (DØ), showing the two-dimensional likelihood and its projections on to each parameter.**

The current status of top mass measurements[10] is shown in Figure 4. The most precise measurements come from the lepton + jets analyses, where the use of $W \to jj$ calibration is an important recent improvement. The figure also shows the indirect constraints on the mass of the Standard Model Higgs that are obtained using this top mass: the best fit is $m_H = 91^{+45}_{-32}$ GeV, and at the 95% CL $m_H < 186$ GeV.

With plausible (but not necessarily easy to achieve) assumptions about evolution of systematic errors, and improvements in b-tagging (using neural networks) and in jet resolution (incorporating tracking information) that are already in progress, we can expect substantial improvements in the precision with which the top mass is measured. An uncertainty of $\Delta m_t = 1.4$ GeV should be attainable with 4 fb$^{-1}$, and 1.2 GeV with 8 fb$^{-1}$ [11].

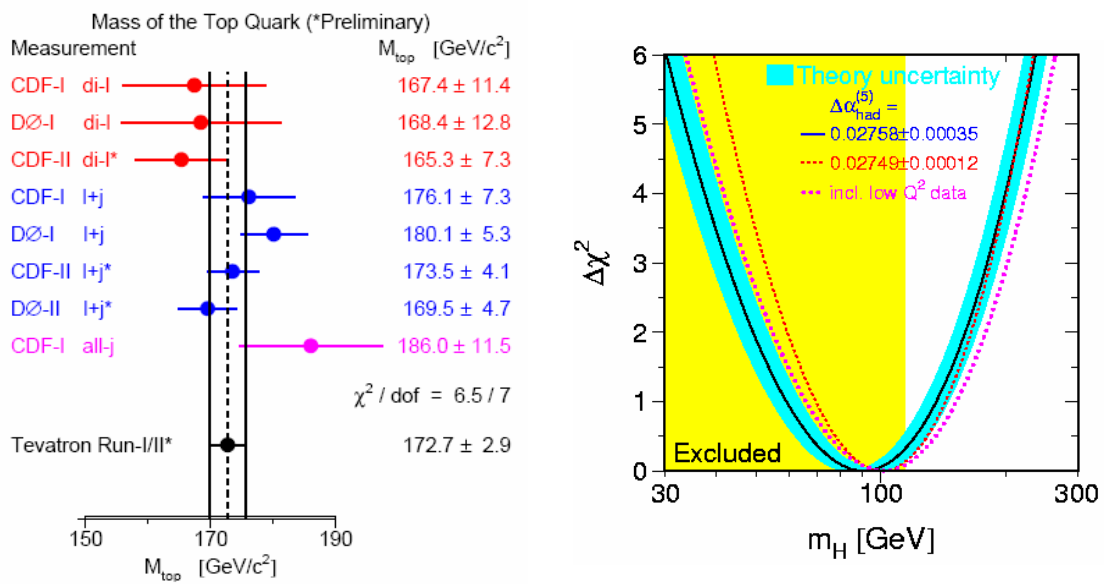

**Figure 4 - Current status of top quark mass measurements and impact on Higgs mass.**

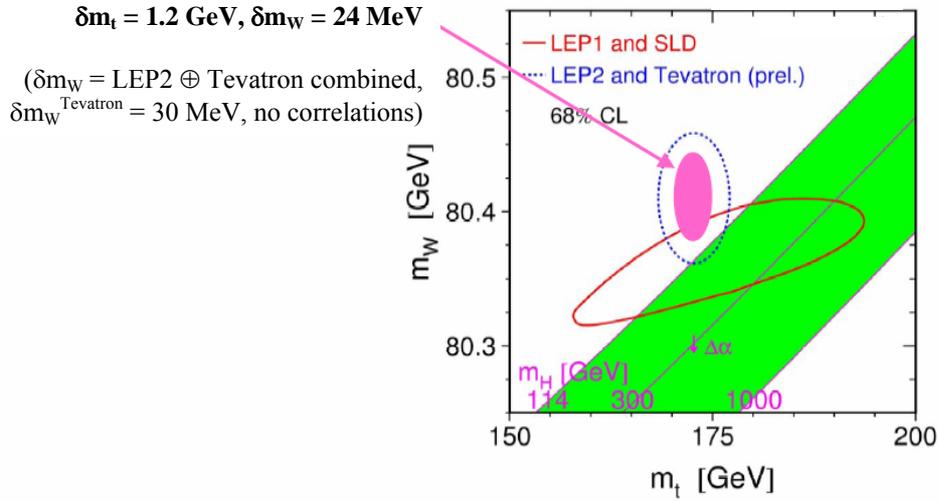

**Figure 5 - A possible scenario for the top mass and W mass at the end of Tevatron Run II.**

To take full advantage of such improvements in the electroweak fits would also require a better W mass measurement. The goal of the Tevatron is to improve on LEP2, which will require an inverse femtobarn or more of data. The general strategy is to extract the W mass from kinematic quantities, usually the transverse mass but also the lepton $p_T$ and missing $E_T$ distributions. The overall scale is set by the Z, using LEP's measurement. CDF carried out an analysis with ~ 200pb$^{-1}$ and obtained an uncertainty of $\Delta m_W$ = 76 MeV (the value of $m_W$ is still blinded). This is consistent with an eventual 30 MeV measurement from the Tevatron, which might give a world average $\Delta m_W$ = 24 MeV. Such a scenario is sketched in Figure 5. This precision yields a 25-30% uncertainty on the Higgs mass [11].

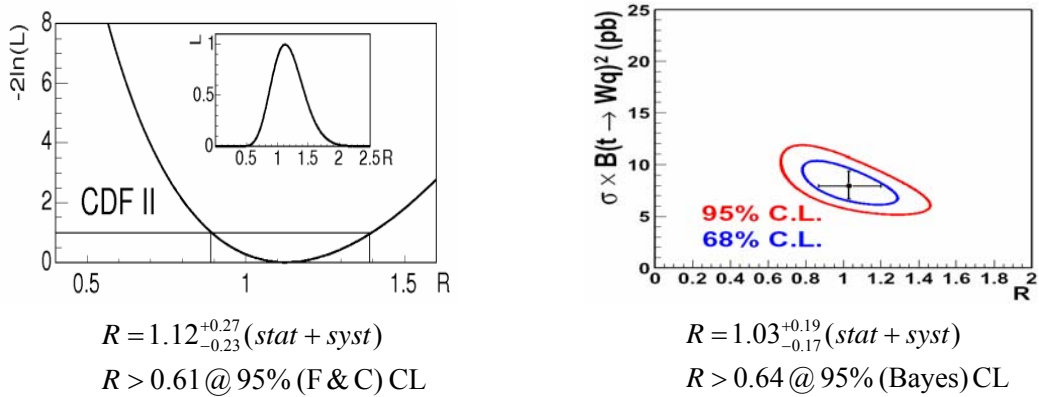

**Figure 6 - Measurements of the ratio R = B(t → b)/B(t → q) from CDF (left) and DØ (right).**

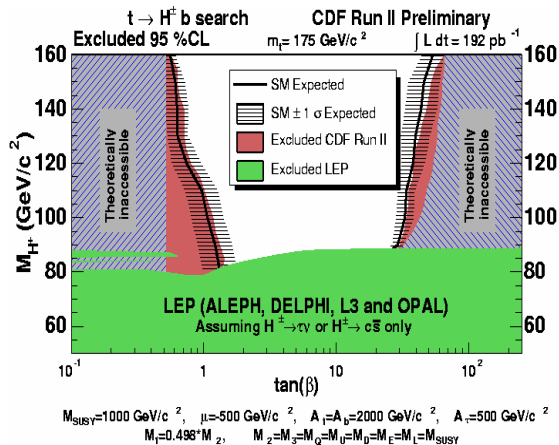
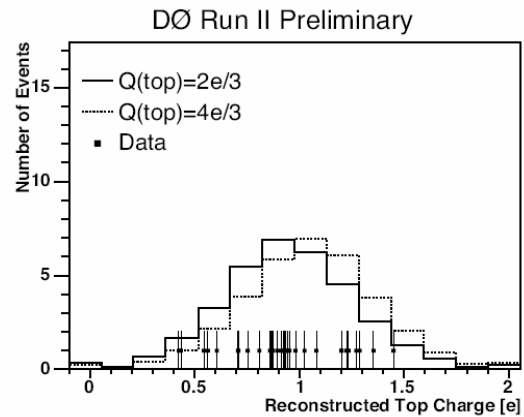

**Figure 7 - Limits on charged Higgs mass as a function of tan β (CDF).**

**Figure 8 - Reconstructed jet charge in top events (DØ).**

**How does top decay?**

In the SM, top decays almost exclusively to a W and a b-quark, but in principle it could decay to other down-type quarks too. The ratio R = B(t → b)/B(t → q) can be extracted[12] by comparing the number of double b-tagged to single b-tagged events. The results (Figure 6) are all consistent with R = 1 as in the standard model, i.e. 100% top → b, but the uncertainties are still quite large.

If there exists a charged Higgs with a mass less than $m_t - m_b$ then the decay t → H$^+$b competes with t → W$^+$b. A sizeable branching ratio B(t → H$^+$b) is expected both at low tanβ where H$^\pm$ → cs and Wbb dominate, and at high tanβ where H$^\pm$ → τν dominates. These decays will have a different effect on cross section measurements in various channels. CDF used $\sigma_{tt}$ measurements in dileptons, lepton+jets and lepton+tau channels, allowed for losses to t → H$^+$b decays, and performed a simultaneous fit to all channels assuming same $\sigma_{tt}$. They are able to exclude some regions of parameter space (see Figure 7) at low and high tanβ, but with the current data there is still room for a substantial branching ratio of t → H$^+$b (as high as 50%, even).

The top quark electric charge has not been measured directly, but it has been possible to distinguish between the hypothesis of (t → W$^+$b) and that of (Q → W$^+$ $\bar{b}$) where Q is an exotic, charge $4/3$ object. Using 21 double-tagged events, DØ finds 17 with a fully convergent kinematic fit; a jet-charge (a $p_T$ weighted sum of track charges) is then constructed for the b-tagged jets. One expects b (q = $-1/3$) to fragment to a jet with leading negative hadrons, but $\bar{b}$ (q = $+1/3$) to fragment to leading positive hadrons. The jet charge distribution in the data (Figure 8) is consistent with a charge $2/3$ top quark and excludes a charge $4/3$ exotic object at the 94% CL.

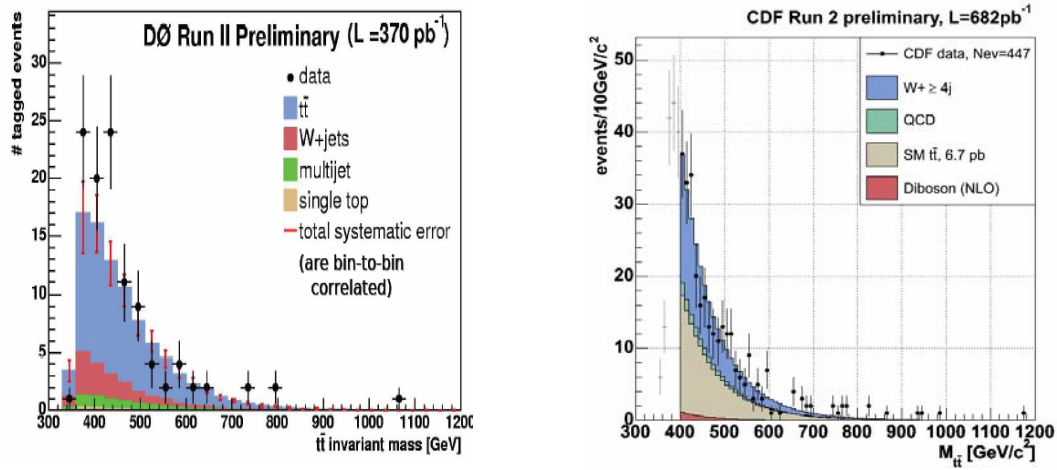

**Figure 9 - Reconstructed $t\bar{t}$ mass distributions from DØ (left) and CDF (right).**

Because its mass is so large, the top quark is expected to decay very rapidly ($10^{-24}$ s) and there is no time to form a top meson. The t → Wb decay then preserves the spin information, which is reflected in the decay angle and momentum of the lepton in the W rest frame. It is found that the fraction of right handed W's is $F_+ < 0.25$ (DØ), 0.27 (CDF) (95%CL), while the fraction of longitudinal W's is $F_0 = 0.74 ^{+0.22}_{-0.34}$ (CDF). In the standard model, $F_+ \approx 0$ and $F_0 \approx 0.7$, so all is consistent.

In principle, the top and antitop spins are correlated in top pair production[12]. DØ carried out an analysis in Run I, using only six events mainly as a proof of principle[14]. CDF have made sensitivity studies concluding that one would need a few fb$^{-1}$ before correlations can be seen. However, at LHC, precision measurements seem possible. One can look at dilepton and l+jets events, in various bases. This may be a useful tool to study or look for nonstandard production mechanisms (such as resonances, or the effects of extra dimensions) or even for CP violation.

**New particles decaying to top?**

One signal might be structure in the $t\bar{t}$ invariant mass distribution from (e.g.) X → $t\bar{t}$. Both experiments have studied the distribution (Figure 9) and both see interesting features, but it is not clear that they are consistent. Since the conference, new CDF results became available[15], and seem more consistent with standard model expectations.

**Single Top Production**

Single top production[16] probes the electroweak properties of top and measures the CKM matrix element |$V_{tb}$|. It is a good place to look for new physics connected with top, and for this

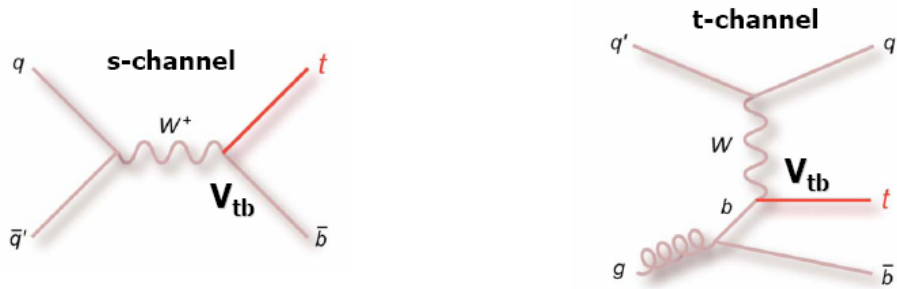

**Figure 10 - Single top production processes.**

reason it is desirable to be able to separate s and t-channel production (Figure 10). The s-channel mode is sensitive to charged resonances, while the t-channel mode is more sensitive to FCNCs and new interactions.

Single top suffers from much higher backgrounds than $t\bar{t}$ production, and even though the cross section is expected to be roughly 1 and 2 pb in the s and t channels respectively, it has still not been observed. The current best limits[17] are from DØ and are $\sigma < 5.0$ and $4.4$ pb in the s and t channels. These are starting to disfavour some models of non-standard physics (Figure 11). The single top search has proved an interesting reference case for advanced analysis techniques. The original analysis was done with simple cuts; moving to a multivariate approach roughly doubled the sensitivity. With the current analysis, a statistically significant observation of single top production will happen in Run II (see Figure 12)[11], but improvements are still desirable. We expect to be able to measure $|V_{tb}|$ to 11% with 4 fb$^{-1}$ and 9% with 8 fb$^{-1}$.

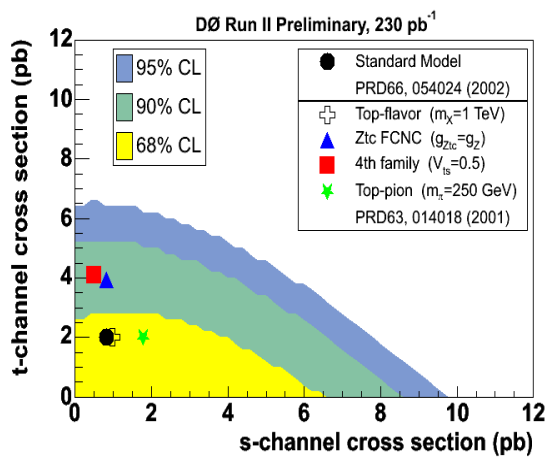
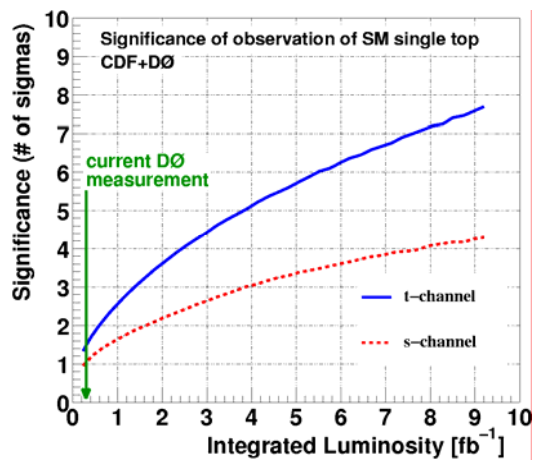

**Figure 11 - DØ limits on single top production in s- and t-channels, together with expectations from the standard model and from a variety of new physics scenarios.**

**Figure 12 - Prospects for single top in Run II.**

**Top at LHC**

The LHC is on track for first collisions in Summer 2007, initial measurements in 2008, and first precision measurements perhaps in 2009 with between 1 and 10 fb$^{-1}$. The LHC has great potential for top physics[18]. The cross section is enormous: one day at the LHC at $10^{33}$ cm$^{-2}$ s$^{-1}$ luminosity is equivalent to ten years at the Tevatron for Standard Model production, because at LHC top is a low-*x*, gluon-dominated process. In many cases, statistical uncertainties will become negligible, and there will be dramatic improvements over statistically limited Tevatron analyses (such as spin, polarisation, and rare decays). This will yield an improved understanding of top and a clearer window on beyond the standard model physics. It is understood that a huge amount of work is needed prior to measurements, to understand the detectors and control systematics. Indeed, early top signals will play an important role in commissioning the detectors, and some of the earliest LHC physics results, and earliest sensitivity to new physics, should come from top physics. The signal is large enough that clear lepton + jets signal can be seen in 150 pb$^{-1}$ with $H_T$ cuts, and no b-tagging required[20]. While top is a background to discovery physics at LHC in modes with leptons, jets and missing $E_T$ (such as H $\rightarrow$ WW, and many supersymmetry final states) it is also a handle for new physics in discovery modes for charged Higgs like tH$^+$ and tbH$^+$ [21]. The $\bar{t}$tH mode[22] also permits a verification of the top Yukawa coupling and hence of the mechanism of fermion mass generation to the 20–30% level.

The enormous statistics available will lead to an even greater emphasis on control of systematics. On the one hand, it will be necessary to worry about issues at < 1% level that are not major concerns at the Tevatron; on the other hand, there will be sufficient statistics to be able to adopt strategies like removing events with identified semileptonic b-decay jets to reduce uncertainties, or to select a high-$p_T$ top sample to reduce combinatorics, if desired.

A number of top mass strategies have been studied[21]. The standard lepton + jets channel will have an excellent signal to background ratio of about 30, and the statistical uncertainty is tiny (100 MeV). It should be possible to push the systematics to the 1 GeV level (the b-jet energy scale is expected to dominate). In the dilepton channel, studies have suggested that the systematics will be at the 1.7 GeV level. Some more exotic possibilities have been studied, such as mass measurements in all-jets channel with a 3 GeV uncertainty, and leptonic final states with J/$\psi$ where the statistics would be low, but one might attain a systematic uncertainty of 0.5 GeV. All these methods have very different sensitivities to systematics, and by combining them it should be possible to measure $m_t$ to ~ 1 GeV with 10 fb$^{-1}$. Such a measurement could be combined with an improved W mass determination, also in principle possible at the LHC, to give a precise consistency test with the mass of the SM Higgs or with the mass spectrum of supersymmetry. Going beyond the LHC, it has been suggested that a measurement at the level of $\Delta m_t$ ~ 100 MeV would be possible at the ILC, but this would require further theoretical progress on higher order calculations[24].

| Electric charge $+2/3$? | Known not to be $4/3$ |
|---|---|
| Colour triplet? | Yes? (production cross section) |
| Spin ½ ? | not really tested – spin correlations at LHC? |
| Isospin ½ ? | Yes? (decays to W and a down-type quark) |
| V – A decay? | Tested at 20% level |
| BR to b quark ~ 100% ? | Tested at 20% level |
| FCNC? | Probed at the 10% level |
| $|V_{tb}|$ | Not yet – will test at 10% level in Run II |
| Top width? | perhaps test with single top? |
| Yukawa coupling? | Not yet – will test at 20-30% level at LHC |
| Top mass | $172.7 \pm 2.9$ GeV |
| Higgs mass | < 186 GeV |

**Table 1 – Current knowledge of Top Quark properties.**

**Single top at LHC**

The cross section for the t-channel process is 120 times higher than at the Tevatron. An ATLAS study[25] concluded that a signal to background ratio of 3 and a statistical uncertainty on the signal of 1.4% could be attained with $30\text{fb}^{-1}$. The s-channel process is harder, since the cross section is only ten times higher than at the Tevatron, but offers the prospect of a direct extraction of $|V_{tb}|$ from the ratio of W* (single top) to real W production. In addition, the tW process (negligible at the Tevatron) should be observable[26]. This is the only single top process where we directly observe the W, and is a more direct measure of top's coupling to W and a down-type quark. The theoretical definition is "delicate" and new work in progress. The major background is from tt; an ATLAS study concluded that while the signal to background ratio is only 1:7, a significant signal could be observed in $30\text{fb}^{-1}$.

**Top spin and polarisation at LHC**

The high statistics available mean that one can significantly improve on the Tevatron[22]. The W helicity in top decay should be measured at the 1 – 7 % level, dominated by systematics. Top spin in single top (expected to be 90% polarised) can be measure at the few % level (at least in fast simulations) and would allow a search for CP violation. Top-antitop spin correlations can be measured at the 10% level, though again this result is based on a fast simulation.

**FCNC decays and HERA**

HERA can be used to set limits on flavour changing neutral current utZ, ut$\gamma$ couplings and on consequent decay branching ratios B(t $\rightarrow$ qZ) and B(t $\rightarrow$ q$\gamma$)[27]. Current limits are at the level of $10^{-1}$–$10^{-2}$, but the LHC will push these to the $10^{-4}$–$10^{-5}$ level. Another example of complementarity with HERA is through the recent (first) experimental determination of the b-quark distribution in the proton, which is needed for calculation of single top production.

**Conclusions**

Table 1 attempts to summarise what we now know about top, and in particular, the extent to which we have tested that is behaves like a standard model up-type quark.

- At the 20% level, top seems to behave like an up type quark which just happens to have an extraordinarily large mass. That mass has been measured very precisely, thus constraining the Higgs sector; but we do not yet know if this mass arises (as in the SM) from a Yukawa coupling or from something more interesting.

- Searches for single top have sufficient sensitivity to see this process soon.

- We are starting to make interesting measurements of spin in top decays, searching for non-standard production and decay mechanisms.

- We have seen fascinating results from the Tevatron with much more to come, and can look forward to a step-change in statistics at LHC, and sensitivity to new observables – perhaps even CP violation in top.